\documentclass[twocolumn,showpacs,amsmath,amssymb,superscriptaddress,nofootinbib]{revtex4}

\usepackage{graphicx}
\usepackage{epsfig}
\usepackage{dcolumn}
\usepackage{bm}
\usepackage{color}

\begin{document}
\title{Triaxiality and shape coexistence in Germanium isotopes}
\author{Lu Guo}
\affiliation {Institut f{\"u}r Theoretische Physik, J. W. Goethe-Universit{\"a}t, D-60438 Frankfurt, Germany}
\author{J. A. Maruhn}
\affiliation {Institut f{\"u}r Theoretische Physik, J. W. Goethe-Universit{\"a}t, D-60438 Frankfurt, Germany}
\author{P.-G. Reinhard}
\affiliation {Institut f{\"u}r Theoretische Physik $\mathit{II}$, Universit{\"a}t Erlangen-N{\"u}rnberg, Staudtstrasse 7, D-91058 Erlangen, Germany}
\date{\today}

\begin{abstract}
The ground-state deformations of the Ge isotopes are investigated in
the framework of Gogny-Hartree-Fock-Bogoliubov (HFB) and Skyrme
Hartree-Fock plus pairing in the BCS approximation. Five different
Skyrme parametrizations are used to explore the influence of different
effective masses and spin-orbit models. There is generally good
agreement for binding energies and deformations (total quadrupole
moment, triaxiality) with experimental data where available (i.e., in
the valley of stability).  All calculations agree in predicting a
strong tendency for triaxial shapes in the Ge isotopes with only a few
exceptions due to neutron (sub-)shell closures.
The frequent occurrence of energetically very close shape isomers indicates
that the underlying deformation energy landscape is very soft.
The general triaxial softness of the Ge isotopes is demonstrated in the fully
triaxial potential energy surface.
The differences between the forces play an increasing role with
increasing neutron number.  This concerns particularly the influence
of the spin-orbit model, which has a visible effect on the trend of
binding energies towards the drip line. Different effective mass plays an important
role in predicting the quadrupole and triaxial deformations.
The pairing strength only weakly affects binding energies and total
quadrupole deformations, but considerably influences triaxiality.
\end{abstract}


\maketitle

\section{\label{level1} Introduction}
In recent years, the development of radioactive nuclear beams and new
gamma-ray detector arrays, together with the powerful
ancillary detectors for light ions, has allowed the experimental
studies of nuclei close to the proton and neutron drip lines
\cite{Simpson99,Mueller99,Tanihata99}, making theoretical studies of
the ground-state properties of these nuclei important. These nuclei
may reveal interesting phenomena of nuclear structure physics and
provide a testing ground for theoretical models, which should explain
the systematics of various properties over long chains of isotopes.

Atomic nuclei exhibit a variety of shapes, varying from spherical to
quadrupole and higher-order multipole deformations.  The possible
shape that the nucleus may adopt results from a delicate balance
between collective and single-particle energies and their dependence
on deformation.  Nuclear triaxiality associated with the breaking of
axial symmetry of the quadrupole deformation, brings up many
interesting collective motions, e.~g., wobbling motion, chiral and
$\gamma$ bands. Recent measurements of the quadrupole moments and
$B(E2)$ transition probabilities from Coulomb excitation experiments
\cite{Sugawara03, Toh00} give more direct indications of nuclear
triaxiality.

Various theoretical methods, e.~g., the shell model
\cite{Hase05,Kaneko02} and self-consistent mean-field models
\cite{Rein99,Bender00} as well as the interacting boson model
\cite{Elliott96}, have been employed to predict nuclear exotic shapes.
Many authors \cite{Teran03,Stoitsov03,Sarri99} discussed the nuclear
shape extensively, but imposed axial and reflection symmetries to
alleviate the complex numerical problems. In axially symmetric
calculations, both prolate and oblate minima with energies very close
to each other can coexist in the nuclear deformation energy curve as a
function of quadrupole deformation.  In such cases, one cannot
definitely conclude which shape is the ground-state configuration and
whether there is a real shape coexistence or if there is yet another
type of deformation, such as $\gamma$-instability with a valley
linking the prolate and oblate shapes through the triaxial region. A
study of the Zn and Ge isotopes within the relativistic mean-field
theory \cite{Gang99} indicates that the restrictions imposed by the
assumption of axial symmetry may be too severe and a triaxial
calculation is necessary.

The motivation for the present work is to study the triaxiality and
shape transitions in a number of Ge isotopes from the proton to the
neutron drip line. These isotopes are the best candidates for
examining the degree to which nuclear ground states can adopt triaxial
deformation, and there are several indications from theoretical
studies already for a potentially rich variation in deformation. For
example, recent HFB calculation predicted that the nucleus
$^{64}\textrm{Ge}$ is extremely soft towards triaxial deformation
\cite{Yamagami01}. Triaxial Routhian surface calculations for
$^{64}\textrm{Ge}$ predicted a well-deformed minimum at $\beta_2=0.3$
and $\gamma\approx 15^\circ$ \cite{Ennis91}.  Calculations using the
IBM-3 approach \cite{Elliott96} described $^{66}\textrm{Ge}$ and
$^{68}\textrm{Ge}$ as vibrational nuclei, and an oblate shape was
predicted for the ground-state band of $^{68}\textrm{Ge}$ by excited
VAMPIR calculations \cite{Petro90}.  Thus, this region should provide
an excellent opportunity to test nuclear structure models.  Moreover,
the study of various isotope chains with different theoretical models
allows the possibility to distinguish what is general and what is
particular in the behavior of these nuclei.

The three most widely used self-consistent mean-field models are: the
Skyrme energy functional, the Gogny force, and a relativistic
mean-field Lagrangian, for a recent review see \cite{Bender03}. In the
present work two mean-field methods are employed to study the nuclear
triaxiality. One is the Skyrme energy density functional plus BCS
pairing, see, e.g., \cite{Bender99}.  In this case it is to be noted
that several parametrizations exist for the Skyrme functional, amongst
which we will choose {five} different ones to cover {a
sufficient} variety of open options.
The other method is the HFB theory with the Gogny interaction
\cite{Gog80,Egido93,Guo04}. The finite-range Gogny force is designed
to provide a simultaneous description of both the particle-hole (HF)
and the particle-particle (pairing) channels of the mean field.

The paper is organized as follows: Section \ref{level2} gives a brief
outline of the two theoretical models, Skyrme HF plus BCS and Gogny
HFB. In Sec.~\ref{level3}, triaxial features of the Ge isotopic chain
are presented within both models. The Skyrme HFBCS calculations
predict shape isomers and $\gamma$-softness in many Ge isotopes.  The
effects of the spin-orbit interaction, the effective mass and the
pairing strength on the systematic properties of the Ge isotopic chain
are discussed with various Skyrme forces and different pairing
strengths. Our calculated ground-state properties are compared with
the available experiments. Section \ref{level4} is devoted to a
summary.

\section{\label{level2} Theoretical Models}

\subsection{Skyrme Hartree-Fock plus BCS}\label{skyrmepar}

The zero-range and density-dependent Skyrme force has been widely
applied to self-consistent nuclear structure calculations owing to its
numerical simplicity. Various parametrizations of the Skyrme force are
available, many of which provide an excellent description of basic
nuclear bulk properties (binding energies, radii), but differ in other
aspects like, e.~g., excitations, fission barriers or neutron matter
properties \cite{Rein06}. For this work we have chosen {five}
typical parametrizations which differ with respect to effective mass
and spin-orbit terms and with respect to the bias in the fit.
{The force SkM$^*$ is meanwhile an established standard, one of
the first forces which managed to cover several observables with
quantitative success \cite{Bar82a}. The force SkT6 stems from a
systematic survey of varied parametrizations at the level of quality
in these years (see SkM$^*$) and it is a choice having an effective
nucleon mass around unity \cite{Tondeur84}.  The force SkP was
originally developed to compute the pairing matrix elements from the
same Skyrme force as used for the mean field \cite{Dob84a}. We take it
here as an alterantive for a force with unit effective mass and employ
the same type of pairing force as for the other parameterizations.
The force SLy6 is a recent fit including more data with emphasis on
isotopic trends, neutron rich nuclei and neutron matter
\cite{Chabanat98}.  It has rather low effective mass.  The force SkI3
is also a recent fit having low effective mass \cite{Rein95}. It employs}
an extended spin-orbit force which
was built in analogy to relativistic models \cite{Fried86}. Its fit
includes also the nuclear charge form factor which, in turn, provides a
more realistic (i.e. softer) surface thickness. 
The effective masses for the {five Skyrme} parametrizations
are given in column two of  Table \ref{table1}.

\begin{table}
\caption{\label{table1} 
The isoscalar effective mass $m^\ast/m$ in infinite nuclear matter and
the pairing strengths $V_p$ for protons and $V_n$ for neutrons
for the Skyrme inqteractions used in the calculations. 
}
\begin{ruledtabular}
\begin{tabular}{cccc}
Force & $m^\ast/m$ & $V_p$ $[\textrm {MeV} \textrm{fm}^3]$ & $V_n$ $[\textrm {MeV} \textrm{fm}^3]$
\\
\hline
SkT6  &  1.00    & -202.526 &  -204.977 \\  
SkP   &  1.00    & -245.540 & -230.997 \\
SkM$^*$ &  0.79  & -279.082 & -258.962 \\
SLy6  &  0.69    & -298.760 &  -288.523 \\
SkI3  &  0.57    & -335.432 &  -331.600 \\
\end{tabular} 
\end{ruledtabular}
\end{table}
The Skyrme energy functional consists of kinetic energy, Skyrme
interaction energy, Coulomb energy including exchange in Slater
approximation, pairing energy, and a correction for the spurious
center-of-mass motion. All terms besides the center-of-mass energy can
be expressed in terms of local distributions, namely density,
kinetic energy density, spin-orbit current, and pair
density.
The pairing correlations are treated in the BCS approximation using a
delta pairing force \cite{Tondeur83,Krieger90},
$V_{\textrm{pair}}(\vec{r},\vec{{r}^\prime})=V_q\delta(\vec{r}-\vec{{r}^\prime})$. The
pairing strength $V_p$ for protons and $V_n$ for neutrons depend on
the mean-field parametrization. For each parametrization they are
separately fitted to the pairing gaps in the selected isotopic and
isotonic chains.  The pairing strengths for the different Skyrme
parametrizations are taken from \cite{Fle04b} and listed in table
\ref{table1}.
{Note that they vary dramatically with the force. They 
depend on the actual shell structure, in particular on 
the effective mass.}
The variation of the energy functional with respect to the
single-particle wave functions yields the mean-field equations and
variation with respect to the occupation amplitudes the associated
pairing equations.

The coupled HF-BCS equations are solved on a grid in coordinate space
with Fourier representation of the derivatives.  No symmetry
restriction has been imposed in the calculation.  The
stationary states are found with the damped gradient iteration method
\cite{Blum92}. As the most sensitive criterion for convergence we take the
variance of the mean-field Hamiltonian which is required to grow
smaller than $10^{-4}$ MeV. Experience shows that this is a sufficient
safeguard against being deceived by pseudo-convergence (an iteration
being stuck at a certain energy value for very long times).
We start the iteration from various initial states to explore the
landscape of isomers which are stable stationary minima besides the
ground state. Position and energy of the isomers help to indicate the
softness of the deformation energy landscape. Note, however, that the
search for isomers is only exploratory. We do not aim at unraveling
the complete landscape of isomers. We rather consider the isomers
which we happen to find as indicators for the softness of the
underlying deformation energy landscape.

\subsection{Gogny Hartree-Fock-Bogoliubov}

We use the finite-range Gogny force with parametrization D1S
\cite{Gog84, Gog91}.  It consists of a finite-range part with Gaussian
shape and containing the four relevant spin-isospin exchanges (Wigner,
Majorana, Heisenberg, and Bartlett), a density-dependent
zero-range term, a zero-range spin-orbit force, and the Coulomb
force. Exchange is treated exactly in all terms.

We have taken into account all the contributions to HF and pairing
fields arising from the Gogny and Coulomb interactions as well as the
two-body correction of the kinetic energy in the self-consistent
procedure.  The finite-range Gogny force allows to derive the HF
Hamiltonian and the pairing field simultaneously from
one and the same Hamiltonian. The practical treatment is, of course,
much more cumbersome. To save CPU time in the numerical calculations,
the $\hat{P}e^{-i\pi\hat{J_z}}$ ($z$-simplex) and
$\hat{P}e^{-i\pi\hat{J_y}}\hat{\tau}$ ($\hat{S_y ^T}$) symmetries are
imposed \cite{JD1,JD2}, where $\hat{P}$ is the parity operator,
$e^{-i\pi\hat{J_i}}$ the rotation operator around the {\it i}-axis
with an angle $\pi$, and $\hat\tau$ the time reversal operator.  Owing
to the $z$-simplex and $\hat{S_y ^T}$ symmetries, a mass asymmetry of
the nucleus is allowed only along the $x$-axis.

The HFB equation has been solved in a three-dimensional harmonic
oscillator basis \cite{Guo04,Guo05,Guo06}. The triaxial oscillator
parameters of the Hermite polynomials were optimized for each nucleus
to obtain the largest ground-state binding energy. In our calculations
of the Ge isotopes, the optimization was done with an expansion
of single-particle wave functions up to a principal quantum number
$N_0=8$ in the oscillator basis. This basis size together with the
optimized basis parameters provides a very good
description, comparable to the calculation with 12 shells in the fixed
basis \cite{Guo04}. No isomeric states could be calculated, however,
because the optimization of the basis produces convergence only into
the ground state.

\section{\label{level3} Triaxiality and Shape Coexistence }

We discuss the properties of ground states and coexistent isomeric
states for the Ge isotopes from the proton-rich isotopes with neutron
number $N=26$ to neutron-rich nuclei until close to the neutron
drip-line at $N=76$.
\begin{figure}
\centerline{\epsfig{figure=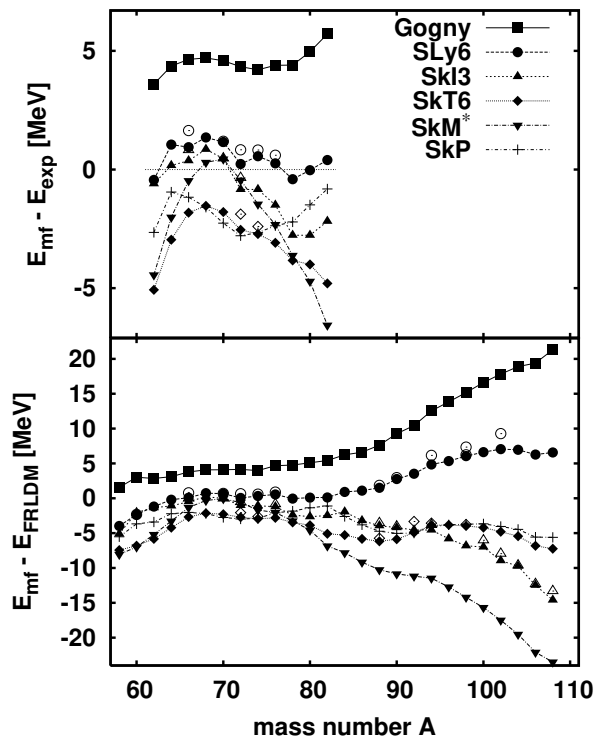,width=8cm}}
\caption{\label{fig:energy-comb} 
Binding energies along the chain of Ge isotopes for the
{six forces Gogny, SkM$^*$, SkT6, SkP, SLy6, and SkI3}, as indicated.
The lower panel shows the energy differences to FRLDM values
\cite{Mol95a}
and the upper panel as differences to the experimental values 
\cite{Audi03}. Filled symbols denote ground states and open 
symbols isomers.
}
\end{figure}
Figure \ref{fig:energy-comb} shows the energies of ground-states
(filled symbols) and isomers (open symbols). The energies as such
would have a huge variation and lead to a rather meaningless plot, so
that we show relative energies, in the upper panel relative to
experiment and in the lower panel relative to the results from a
macroscopic-microscopic model, the finite-range liquid-drop model
(FRLDM) with results taken from \cite{Mol95a}.
In the range of experimentally accessible isotopes, there is a clear
sorting of energies with forces, Gogny predicts
underestimated binding 
{throughout.
The older Skyrme forces (SkM$^*$, SkT6, SkP) tend to overbinding.
}
The typical deviation from experiment is smallest for SLy6 and SkI3
which is no surprise because these two forces are the most recent
developments in our sample.
The exotic regime 
{(i.e. the wings of the distributions)}
shows a much wider span of energy predictions, as is
expected since the uncertainty in extrapolations grows when going
farther away from the regime of stable nuclei for which the forces were 
adjusted. 
{Even the ordering of energies with forces
is interchanged which is due to different isovector properties in the
different parameterizations.}

{The lower part of the figure shows the energies in a much broader
range of isotopes drawn as difference to FRLDM values because
experimental data are not available in that deeply exotic regime.  The
span of the predictions grows substantially as could be expected in a
regime of bold extrapolation. The trends which had already started to
develop in the range of experimentally accessible nuclei (see upper
panel) are basically continued with Gogny tending to lowest binding
energies while the Skyrme forces span a broad band up to very strong
binding at large neutron excess for SkM$^*$.  Simple dependences, 
e.g. on the effective mass, cannot be seen.  There is a mix of various
influences, symmetry energy and shell structure determined by
effective mass as well as spin-orbit splitting. Note that the trend of
SkI3 to stronger binding develops late on the isotopic chain, see the
crossing of trends at the rather large $A=94$.  That is a consequence
of the different isotopic mix in the spin-orbit term whose impact
grows with neutron excess \cite{Rein95}.  }
%
%

%
\begin{figure}
\centerline{\epsfig{figure=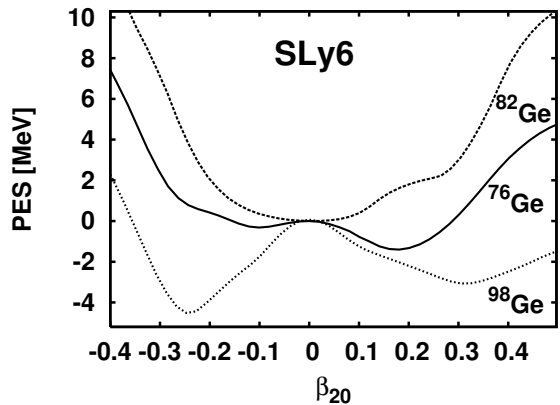,width=8cm}}
\caption{\label{fig:Ge_pes_sly6} 
Potential-energy surfaces (PES) versus axially symmetric
quadrupole deformation computed with SLy6 for three selected 
Ge isotopes as indicated. Positive quadrupole moments
correspond to prolate shapes and negative moments to oblate ones.
}
\end{figure}
\begin{figure}
\centerline{\epsfig{figure=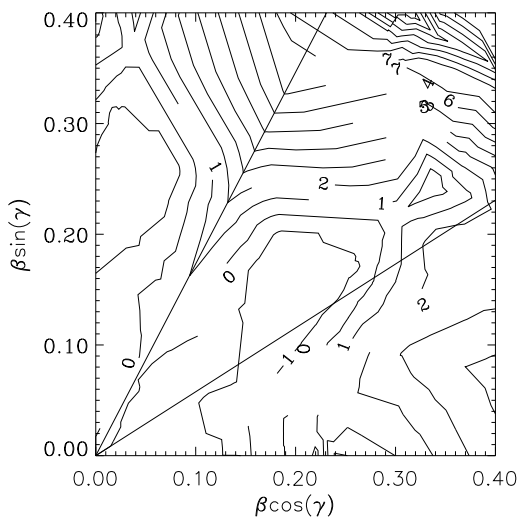,width=6.8cm}}
\caption{\label{fig:Ge_tri_sly6} 
Triaxial PES with SLy6 for the nucleus $^{76}\textrm{Ge}$.
The contour lines are labelled with the energy in MeV.
The two straight lines indicate the directions of triaxial 
deformation $\gamma=30^\circ$ and $60^\circ$.}
\end{figure}
For the Skyrme energy functional there appear isomers in many cases
and it is noteworthy that isomer energies are usually extremely
close. This indicates that the energy landscape is very soft, often
producing energetically competing shapes.  The dominant shape
parameter is the quadrupole deformation $Q_{2\mu}=\langle
r^2Y_{2\mu}\rangle$. For better comparison, it is useful to handle it
in terms of the dimensionless quadrupole deformations
\begin{subequations}
\label{eq:deform}
\begin{eqnarray}
  \beta
  &=&
  \sqrt{\sum_\mu\beta_{2\mu}^2}
  \quad,
\\
  \gamma
  &=&
  \mbox{atan}\Big(\frac{\beta_{22}+\beta_{2-2}}{\sqrt{2}\beta_{20}}\Big)
  \quad,
\\
  \beta_{2\mu}
  &=&
  \frac{4\pi\langle r^2Y_{2\mu}\rangle}{5\langle r^2\rangle}
  \quad.
\end{eqnarray}
\end{subequations}
Figure \ref{fig:Ge_pes_sly6} tries to visualize the typical energy
landscapes by showing a cut of the potential-energy surfaces (PES)
along axial shapes (i.e. along $\beta_{20}$) for three examples, a
proton-rich isotope $^{76}$Ge, the most stable nucleus $^{82}$Ge,
and $^{98}$Ge as a very neutron-rich case. The PES are indeed very
soft, particularly for $^{76}$Ge and $^{82}$Ge. More structures seem
to develop on the neutron-rich side.  However, the prolate and oblate
minima are softly connected along the path of triaxiality having, in
fact, a shallow minimum at a fully triaxial shape (see table
\ref{tab3} later on). 
The general triaxial softness of the Ge isotopes is demonstrated 
for the case of  $^{76}$Ge in 
Figure \ref{fig:Ge_tri_sly6}, which shows the PES in the fully triaxial
deformation landscape.
The weak prolate minimum extends, in fact, as a shallow valley deep
into the regime of triaxiality.
In this way the PES will be very soft in many
cases, with the actual ground state having large shape fluctuations
about the minima in the mean-field configurations.  This requires, in
principle, the computation of the collective ground-state correlations
as done, e.g., in the surveys of \cite{Fle04a,Ben05a,Ben06a}.
Such calculations, however, are very cumbersome and still restricted to
axial symmetry. For a first exploration, the
deformations at the mean-field minima still provide useful
guidelines for the structure and low-energy dynamics of the nuclei.

\begin{figure*}
\centerline{\epsfig{figure=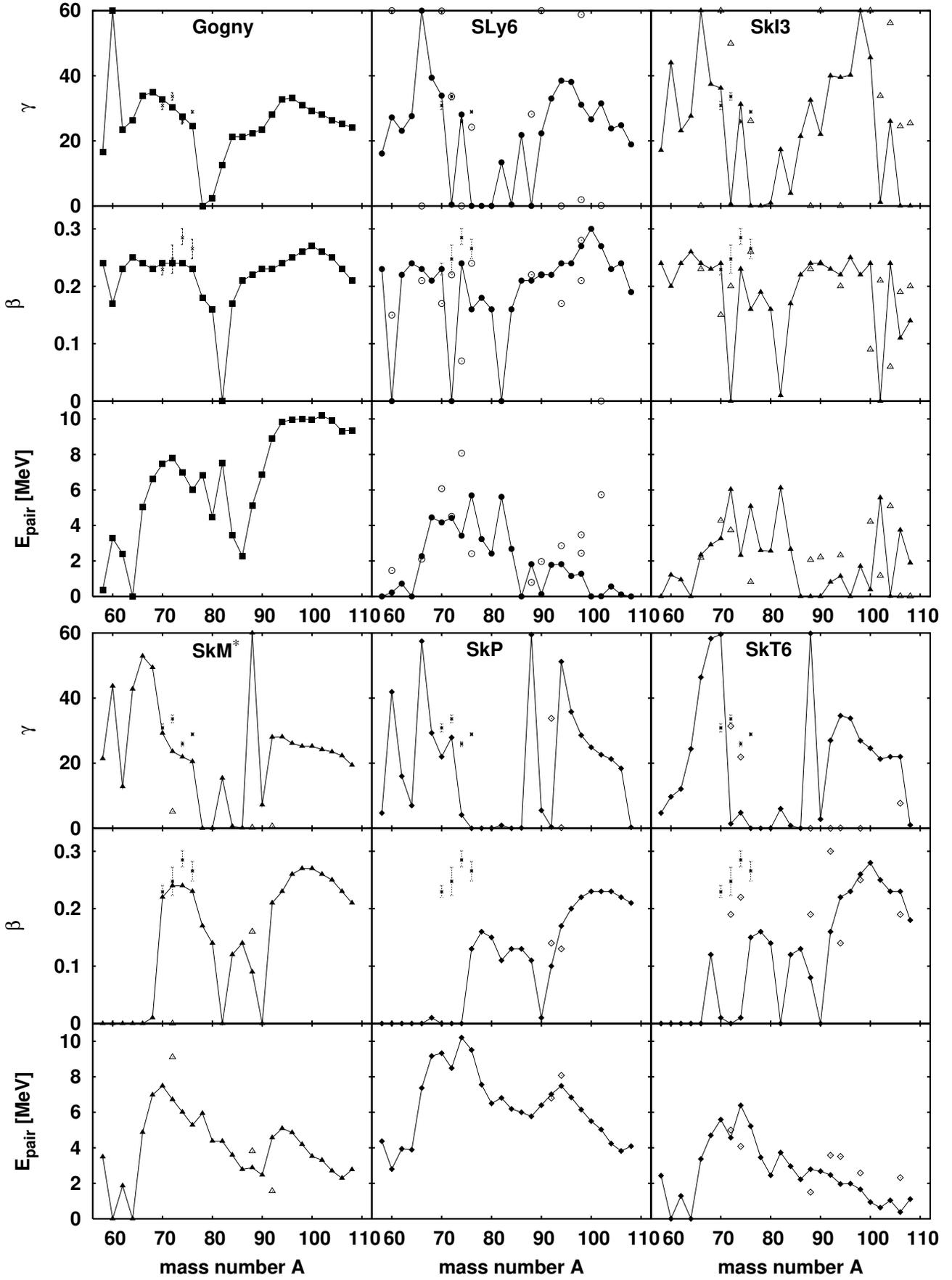,width=16.8cm}}
\caption{\label{fig:shapes2} 
The shape parameters $\beta$ for total quadrupole deformation
(middle panels) as well as $\gamma$ for the triaxiality (upper panels),
and the pairing energy (lower panels) for the
{six forces: Gogny (upper left panels), 
SLy6 (upper middle), SkI3 (upper right), 
SkM$^*$ (lower left), SkP (lower middle),
and SkT6 (lower right).}
The ground-state values are shown with full symbols. Isomers 
are indicated by open symbols.
Available experimental values for deformation and triaxiality
are indicated by error bars  \cite{Sugawara03, Toh00}.
}
\end{figure*}

Figure \ref{fig:shapes2} summarizes the results for deformations and
pairing energies. The quadrupole deformation is quantified in terms of
$\beta$ and triaxiality $\gamma$ as defined in
Eq. (\ref{eq:deform}). Tables \ref{tab1}, \ref{tab2}, and \ref{tab3}
supply the complementary detailed numbers.
All calculations agree in predicting a strong tendency to triaxial
shapes for the Ge isotopes and they all, except SkP, show a clearly spherical
$^{82}\textrm{Ge}$, which is no surprise because neutron number $N=50$
corresponds to a closed neutron shell.
Previous theoretical works \cite{Yamagami01,Ennis91} have selected a
few specific nuclei in this mass region to study nuclear exotic
shapes, thus motivating the present systematic study of the isotopic
chains from the proton to the neutron drip-line. For example, HFB in
coordinate space and the two-basis method \cite{Yamagami01} predicted
the ground-state deformation of the nucleus $^{64}\textrm {Ge}$ as
$\beta=0.27$ and $\gamma= 25^\circ$, and triaxial Routhian surface
calculations for $^{64}\textrm{Ge}$ yielded a well-deformed ground
state at $\beta_2=0.3$ and $\gamma\approx 15^\circ$
\cite{Ennis91}. They predicted similar ground-state deformations for
$^{64}\textrm {Ge}$ as our Gogny, SLy6, and SkI3 calculations (as
shown in Tab.~\ref{tab1}). {At variance are the predictions
of SkM$^*$, SkP, and SkT6 which have the higher effective masses in
our sample and which all yield a spherical ground state.
This isotope with 32 neutrons is in the transitional region
still near the magic 28 and such nuclei are sensitive to faint
changes in shell structure.}

At second glance, we see {more} interesting differences in detail.  The
simplest pattern is provided by the Gogny force.  Except the closed shell
$^{82}\textrm {Ge}$, all nuclei are well-deformed and all
of these except two are clearly triaxial.  Quite similar global
deformation is found also for SLy6 and SkI3. Here we have a few more
spherical ground states related to the known neutron sub-shell closures
($N=40$ and $N=28$ for SLy6, $N=40$ and $N=70$ for SkI3).  
More
differences are seen concerning triaxiality. While the Gogny force
predicts triaxial minima almost throughout, SLy6 and SkI3 have many
axially symmetric exceptions near the stable isotopes. In these cases
there are, however, many triaxial isomers 
and the energy separations are generally very small
which indicates that the
situation is extremely soft in triaxial direction. A certain ambiguity
between axial and triaxial deformations is also apparent from the
zigzag pattern of triaxiality along the isotopic chain. Fully
correlated calculations would smooth the fluctuations and, seen with a
smoothing filter, the predictions of Gogny, SLy6 and SkI3 are in
general quite similar, predicting the general importance of
triaxiality for the Ge isotopes.

The situation is somewhat different for 
the other three Skyrme forces shown in the lower panels. These
forces agree
nicely with the others for the very neutron-rich isotopes but yield
generally smaller deformations near the valley of stability. This is
related to the higher effective mass which acts here to reduce the
energy gain from deformation for the mid-shell neutron numbers.
{The three forces SkM$^*$, SkP, and SkT6 have a broad transitional
region of sphericity around the magic $N=28$ before deformation
develops.  Further spherical dips appear at the (sub-)shell closures
$N=50$ and $N=58$.  Both features are caused by shell structure,
i.e. the higher effective mass of these three forces.  The majority of
isotopes carrying deformation are similar, although the two forces SkP
and SkT6 with effective mass one have generally smaller deformations.
They are also distinguished by showing broad regions of axiality
around $A=80$ where the other forces show more transitional behaviour,
switching between axial and triaxial shapes. 
In spite of all these differences in detail, we have to keep in mind
that all forces agree in predicting softness and triaxiality in the
region of neutron rich Ge isotopes.
}

The experimental information on the ground-state shapes of the nuclei
$^{70-76}\textrm {Ge}$ \cite{Sugawara03, Toh00} is shown for
comparison. There is good agreement for Gogny, SLy6, {SkI3 and
SkM$^*$}.  The data seems to contradict the sub-shell closure at
$N=40$ seen for SLy6 and SkI3. But note that we see pronounced isomers
in these cases, which indicates a great openness to triaxial effects
when correlations are taken into account. Significant differences are
seen for SkT6 {and SkP}, although also here the isomers provide a
more balanced view. Note that {these two forces with effective
mass one} produce a particularly soft PES and the pure mean-field
predictions become somewhat unreliable here.

\begin{table}
\caption{\label{tab1} The properties of ground states (GS) and
coexistent isomeric states (IS) for Ge isotopes with the Gogny HFB and
Skyrme HF-BCS using the SLy6, SkI3 and SkT6 parametrizations.  The
eight columns represent the nucleus calculated, the force used, the
type of state, total, HF and pairing energies, quadrupole deformation
$\beta$, and triaxial deformation $\gamma$ (given in degrees). The
available experimental results are listed for comparison.}
\begin{ruledtabular}
\begin{tabular}{cccccccc}
nucleus & force & state &$E_B$[MeV] & $E_{hf}$ & $E_{pair}$ & $\beta$ & $\gamma$ \\
\hline
$^{58}{\textrm {Ge}}$  &  Gogny & GS  & -450.30  & -449.94   &   0.36   &  0.24   &  16.6  \\
                       &  SLy6  & GS  & -455.83  & -455.82   &   0.00   &  0.23   &  16.1 \\ 
                       &  SkI3  & GS  & -456.97  & -456.97   &   0.00   &  0.24   &  17.1 \\ 
                       &  SkT6  & GS  & -459.30  & -456.87   &   2.43   &  0.00   &   4.7 \\
                     &SkM$^\ast$& GS  & -459.91  & -456.42   &   3.49   &  0.00   &  21.4  \\ 
                       &  SkP   & GS  & -457.39  & -453.01   &   4.37   &  0.00   &   4.7 \\
\hline
$^{60}{\textrm {Ge}}$  &  Gogny & GS  & -483.87  & -480.60   &   3.27   &  0.17   &  60.0  \\
                       &  SLy6  & GS  & -489.23  & -489.02   &   0.22   &  0.00   &  27.2 \\ 
                       &        & IS  & -489.21  & -487.75   &   1.46   &  0.15   &  60.0 \\
                       &  SkI3  & GS  & -488.92  & -487.71   &   1.21   &  0.20   &  44.0 \\ 
                       &  SkT6  & GS  & -493.70  & -493.70   &   0.00   &  0.00   &   9.7 \\
                     &SkM$^\ast$& GS  & -493.89  & -493.89   &   0.00   &  0.00   &  43.7 \\
                       &  SkP   & GS  & -490.56  & -487.75   &   2.80   &  0.00   &  41.9 \\
\hline
$^{62}{\textrm {Ge}}$  &  Gogny & GS  & -514.03  & -511.63   &   2.40   &  0.23   &  23.5  \\
                       &  SLy6  & GS  & -518.07  & -517.35   &   0.72   &  0.22   &  23.1 \\
                       &  SkI3  & GS  & -518.23  & -517.29   &   0.94   &  0.24   &  23.1 \\ 
                       &  SkT6  & GS  & -522.70  & -521.41   &   1.29   &  0.00   &  12.1 \\
                     &SkM$^\ast$& GS  & -522.07  & -520.21   &   1.86   &  0.00   &  12.8 \\
                       &  SkP   & GS  & -520.28  & -516.34   &   3.94   &  0.00   &  16.0 \\
                       &  Exp.  & GS  & -517.63	 &           &          &         &       \\
\hline
$^{64}{\textrm {Ge}}$  &  Gogny & GS  & -541.52  & -541.52   &   0.00   &  0.25   &  26.4  \\
                       &  SLy6  & GS  & -544.83  & -544.83   &   0.00   &  0.24   &  27.6 \\
                       &  SkI3  & GS  & -545.71  & -545.71   &   0.00   &  0.26   &  27.6 \\ 
                       &  SkT6  & GS  & -548.84  & -548.84   &   0.00   &  0.00   &  24.4 \\
                     &SkM$^\ast$& GS  & -547.89  & -547.89   &   0.00   &  0.00   &  42.8 \\
                       &  SkP   & GS  & -546.83  & -542.95   &   3.89   &  0.00   &   7.0 \\ 
                       &  Exp.  & GS  & -545.88	 &           &          &         &       \\
\hline
$^{66}{\textrm {Ge}}$  &  Gogny & GS  & -564.65  & -559.63   &   5.02   &  0.24   &  33.8  \\
                       &  SLy6  & GS  & -568.35  & -566.08   &   2.27   &  0.23   &  60.0 \\
                       &        & IS  & -567.65  & -565.55   &   2.10   &  0.21   &   0.0 \\
                       &  SkI3  & GS  & -568.92  & -566.57   &   2.34   &  0.24   &  60.0 \\ 
                       &        & IS  & -568.47  & -566.29   &   2.18   &  0.23   &   0.0 \\ 
                       &  SkT6  & GS  & -571.11  & -567.75   &   3.37   &  0.00   &  46.4 \\
                     &SkM$^\ast$& GS  & -569.76  & -564.89   &   4.87   &  0.00   &  52.9 \\  
                       &  SkP   & GS  & -570.46  & -563.08   &   7.37   &  0.00   &  57.5 \\
                       &  Exp.  & GS  & -569.29	 &           &          &         &       \\
\hline
$^{68}{\textrm {Ge}}$  &  Gogny & GS  & -586.09  & -579.47   &   6.62   &  0.23   &  35.0  \\
                       &  SLy6  & GS  & -589.44  & -584.99   &   4.45   &  0.21   &  39.4 \\
                       &  SkI3  & GS  & -589.94  & -587.03   &   2.91   &  0.23   &  37.4 \\
                       &  SkT6  & GS  & -592.32  & -587.62   &   4.70   &  0.12   &  58.3 \\
                     &SkM$^\ast$& GS  & -590.49  & -583.51   &   6.98   &  0.01   &  49.4 \\ 
                       &  SkP   & GS  & -592.38  & -583.20   &   9.17   &  0.01   &  29.3 \\ 
                       &  Exp.  & GS  & -590.79	 &           &          &         &       \\
\hline
$^{70}{\textrm {Ge}}$  &  Gogny & GS  & -605.94  & -598.47   &   7.47   &  0.24   &  32.7  \\
                       &  SLy6  & GS  & -609.36  & -605.19   &   4.17   &  0.23   &  33.9 \\
                       &        & IS  & -609.33  & -603.26   &   6.07   &  0.17   &  59.9 \\
                       &  SkI3  & GS  & -610.06  & -606.80   &   3.26   &  0.24   &  36.2 \\
                       &        & IS  & -610.05  & -605.78   &   4.27   &  0.15   &  60.0 \\ 
                       &  SkT6  & GS  & -612.31  & -606.72   &   5.59   &  0.01   &  59.6 \\
                     &SkM$^\ast$& GS  & -610.08  & -602.59   &   7.49   &  0.22   &  29.2 \\ 
                       &  SkP   & GS  & -612.78  & -603.45   &   9.33   &  0.00   &  22.0 \\
                       &  Exp.  & GS  & -610.52	 &           &          &  0.23   &  30.9 \\
\end{tabular}
\end{ruledtabular}
\end{table}

\begin{table}
\caption{\label{tab2}(continued) as table~\ref{tab1}.}
\begin{ruledtabular}
\begin{tabular}{cccccccc}
nucleus & force & state &$E_B$[MeV] & $E_{hf}$ & $E_{pair}$ & $\beta$ & $\gamma$  \\
\hline
$^{72}{\textrm {Ge}}$  &  Gogny & GS  & -624.35  & -616.54   &   7.81   &  0.24   &  30.4  \\
                       &  SLy6  & GS  & -628.45  & -624.03   &   4.42   &  0.00   &   0.4 \\
                       &        & IS  & -627.85  & -623.33   &   4.51   &  0.22   &  33.6 \\
                       &  SkI3  & GS  & -629.52  & -623.49   &   6.03   &  0.00   &   0.6 \\ 
                       &        & IS  & -629.03  & -625.30   &   3.73   &  0.20   &  49.9 \\ 
                       &  SkT6  & GS  & -631.22  & -626.66   &   4.57   &  0.00   &   1.4 \\
                       &        & IS  & -630.56  & -625.56   &   5.00   &  0.19   &  31.4 \\
                     &SkM$^\ast$& GS  & -629.13  & -622.41   &   6.72   &  0.24   &  23.6 \\
                       &        & IS  & -628.90  & -619.79   &   9.11   &  0.00   &   5.1 \\
                       &  SkP   & GS  & -631.47  & -622.98   &   8.49   &  0.00   &  27.9 \\
                       &  Exp.  & GS  & -628.68	 &           &          &  0.25   &  33.7 \\
\hline
$^{74}{\textrm {Ge}}$  &  Gogny & GS  & -641.46  & -634.46   &   7.00   &  0.24   &  27.5  \\
                       &  SLy6  & GS  & -645.10  & -641.67   &   3.42   &  0.24   &  28.1 \\
                       &        & IS  & -644.84  & -636.77   &   8.07   &  0.07   &   0.0 \\
                       &  SkI3  & GS  & -646.50  & -644.18   &   2.32   &  0.23   &  31.2 \\ 
                       &  SkT6  & GS  & -648.38  & -641.99   &   6.39   &  0.01   &   4.8 \\
                       &        & IS  & -648.06  & -643.98   &   4.08   &  0.22   &  21.9 \\
                     &SkM$^\ast$& GS  & -647.12  & -641.11   &   6.01   &  0.24   &  21.9 \\
                       &  SkP   & GS  & -648.32  & -638.11   &   10.21  &  0.00   &   4.1 \\
                       &  Exp.  & GS  & -645.66	 &           &          &  0.28   &  25.9 \\
\hline
$^{76}{\textrm {Ge}}$  &  Gogny & GS  & -657.21  & -651.21   &   6.00   &  0.23   &  24.6  \\
                       &  SLy6  & GS  & -661.34  & -655.65   &   5.69   &  0.16   &   0.0 \\
                       &        & IS  & -661.00  & -658.59   &   2.41   &  0.24   &  24.2 \\
                       &  SkI3  & GS  & -663.10  & -658.01   &   5.08   &  0.16   &   0.0 \\ 
                       &        & IS  & -662.72  & -661.91   &   0.81   &  0.26   &  26.1 \\ 
                       &  SkT6  & GS  & -664.69  & -659.47   &   5.22   &  0.15   &   0.0 \\
                     &SkM$^\ast$& GS  & -663.92  & -658.64   &   5.28   &  0.23   &  20.5 \\ 
                       &  SkP   & GS  & -663.97  & -654.45   &   9.51   &  0.13   &   0.0 \\
                       &  Exp.  & GS  & -661.60	 &           &          &  0.27   &  28.9 \\
\hline
$^{78}{\textrm {Ge}}$  &  Gogny & GS  & -671.98  & -665.14   &   6.84   &  0.18   &   0.0  \\
                       &  SLy6  & GS  & -676.79  & -673.56   &   3.23   &  0.18   &   0.0 \\
                       &  SkI3  & GS  & -679.15  & -676.56   &   2.59   &  0.19   &   0.0 \\ 
                       &  SkT6  & GS  & -680.21  & -676.76   &   3.46   &  0.16   &   0.0 \\
                     &SkM$^\ast$& GS  & -680.00  & -674.06   &   5.94   &  0.17   &   0.0 \\ 
                       &  SkP   & GS  & -678.59  & -671.03   &   7.56   &  0.16   &   0.0 \\
                       &  Exp.  & GS  & -676.38	 &           &          &         &       \\
\hline
$^{80}{\textrm {Ge}}$  &  Gogny & GS  & -685.22  & -680.74   &   4.48   &  0.16   &   2.3  \\
                       &  SLy6  & GS  & -690.21  & -687.79   &   2.42   &  0.16   &   0.0 \\
                       &  SkI3  & GS  & -692.95  & -690.39   &   2.56   &  0.16   &   0.9 \\
                       &  SkT6  & GS  & -694.18  & -691.73   &   2.45   &  0.14   &   0.0 \\
                     &SkM$^\ast$& GS  & -694.89  & -690.51   &   4.38   &  0.14   &   0.0 \\
                       &  SkP   & GS  & -691.67  & -685.17   &   6.50   &  0.15   &   0.0 \\
                       &  Exp.  & GS  & -690.18	 &           &          &         &       \\
\hline
$^{82}{\textrm {Ge}}$  &  Gogny & GS  & -696.72  & -689.21   &   7.51   &  0.00   &  12.6  \\
                       &  SLy6  & GS  & -702.03  & -696.42   &   5.61   &  0.00   &  13.4 \\
                       &  SkI3  & GS  & -704.61  & -698.48   &   6.12   &  0.01   &  17.3 \\
                       &  SkT6  & GS  & -707.23  & -703.50   &   3.73   &  0.00   &   6.0  \\
                     &SkM$^\ast$& GS  & -708.99  & -704.62   &   4.37   &  0.00   &  15.4  \\ 
                       &  SkP   & GS  & -703.25  & -696.44   &   6.81   &  0.11   &   0.9 \\
                       &  Exp.  & GS  & -702.43	 &           &          &         &        \\
\hline
$^{84}{\textrm {Ge}}$  &  Gogny & GS  & -704.51  & -701.07   &   3.44   &  0.17   &  21.2  \\
                       &  SLy6  & GS  & -709.84  & -707.16   &   2.68   &  0.16   &   0.4 \\  
                       &  SkI3  & GS  & -712.69  & -710.02   &   2.66   &  0.17   &   3.9 \\ 
                       &  SkT6  & GS  & -716.02  & -713.06   &   2.96   &  0.12   &   0.8  \\
                     &SkM$^\ast$& GS  & -718.57  & -714.99   &   3.59   &  0.12   &  0.6  \\
                       &  SkP   & GS  & -713.36  & -707.17   &   6.19   &  0.13   &  0.0  \\ 
\hline
$^{86}{\textrm {Ge}}$  &  Gogny & GS  & -711.45  & -709.18   &   2.27   &  0.21   &  21.2  \\
                       &  SLy6  & GS  & -716.96  & -716.96   &   0.00   &  0.21   &  21.8 \\ 
                       &  SkI3  & GS  & -721.26  & -721.26   &   0.00   &  0.22   &  21.4 \\ 
                       &  SkT6  & GS  & -723.88  & -721.66   &   2.22   &  0.13   &   0.0  \\
                     &SkM$^\ast$& GS  & -727.23  & -724.45   &   2.78   &  0.14   &   0.1  \\
                       &  SkP   & GS  & -721.87  & -715.86   &   6.00   &  0.13   &   0.1 \\
\end{tabular}
\end{ruledtabular}
\end{table}

\begin{table}
\caption{\label{tab3}(continued) as table~\ref{tab2}.}
\begin{ruledtabular}
\begin{tabular}{cccccccc}
nucleus & force & state &$E_B$[MeV] & $E_{hf}$ & $E_{pair}$ & $\beta$ & $\gamma$ \\
\hline
$^{88}{\textrm {Ge}}$  &  Gogny & GS  & -716.57  & -711.45   &   5.12   &  0.22   &  22.4  \\
                       &  SLy6  & GS  & -722.69  & -720.87   &   1.82   &  0.21   &   0.0 \\
                       &        & IS  & -722.34  & -721.55   &   0.79   &  0.22   &  28.2 \\
                       &  SkI3  & GS  & -728.13  & -728.13   &   0.00   &  0.24   &  32.5 \\ 
                       &        & IS  & -727.75  & -725.69   &   2.06   &  0.23   &   0.0 \\ 
                       &  SkT6  & GS  & -730.37  & -727.59   &   2.79   &  0.08   &  59.9 \\
                       &        & IS  & -729.92  & -728.43   &   1.50   &  0.19   &   0.0 \\
                     &SkM$^\ast$& GS  & -734.45  & -731.58   &   2.88   &  0.09   &  60.0 \\
                       &        & IS  & -734.01  & -730.20   &   3.81   &  0.16   &   0.2 \\
                       &  SkP   & GS  & -728.92  & -723.15   &   5.77   &  0.11   &  59.5 \\
\hline
$^{90}{\textrm {Ge}}$  &  Gogny & GS  & -720.58  & -713.71   &   6.87   &  0.23   &  23.5  \\
                       &  SLy6  & GS  & -727.08  & -726.95   &   0.13   &  0.22   &  22.3 \\
                       &        & IS  & -726.88  & -724.91   &   1.97   &  0.22   &  60.0 \\
                       &  SkI3  & GS  & -734.05  & -734.05   &   0.00   &  0.24   &  22.0 \\ 
                       &        & IS  & -733.88  & -731.67   &   2.21   &  0.24   &  60.0 \\ 
                       &  SkT6  & GS  & -735.72  & -733.03   &   2.68   &  0.00   &   2.8  \\
                     &SkM$^\ast$& GS  & -740.70  & -738.23   &   2.47   &  0.00   &   7.2  \\
                       &  SkP   & GS  & -734.88  & -728.49   &   6.40   &  0.01   &   5.5 \\ 
\hline
$^{92}{\textrm {Ge}}$  &  Gogny & GS  & -723.74  & -714.83   &   8.91   &  0.23   &  28.1 \\
                       &  SLy6  & GS  & -730.65  & -728.87   &   1.78   &  0.22   &  33.0 \\
                       &  SkI3  & GS  & -738.72  & -737.90   &   0.81   &  0.23   &  40.0 \\
                       &  SkT6  & GS  & -739.11  & -736.64   &   2.47   &  0.16   &  27.0 \\
                       &        & IS  & -737.51  & -733.93   &   3.58   &  0.30   &   0.0 \\
                     &SkM$^\ast$& GS  & -745.35  & -740.78   &   4.57   &  0.21   &  28.0 \\
                       &        & IS  & -744.03  & -742.46   &   1.56   &  0.35   &   0.6 \\
                       &  SkP   & GS  & -739.17  & -732.15   &   7.02   &  0.10   &   0.4 \\
                       &        & IS  & -739.07  & -732.27   &  6.80    &  0.14   &  33.8 \\
\hline
$^{94}{\textrm {Ge}}$  &  Gogny & GS  & -726.31  & -716.48   &   9.83   &  0.24   &  32.7  \\
                       &  SLy6  & GS  & -734.00  & -732.18   &   1.82   &  0.24   &  38.5 \\
                       &        & IS  & -732.69  & -729.83   &   2.86   &  0.17   &   0.0 \\ 
                       &  SkI3  & GS  & -743.40  & -742.26   &   1.14   &  0.22   &  39.5 \\
                       &        & IS  & -742.67  & -740.36   &   2.31   &  0.20   &   0.0 \\ 
                       &  SkT6  & GS  & -742.74  & -740.78   &   1.96   &  0.22   &  34.6  \\
                       &        & IS  & -742.41  & -738.90   &   3.51   &  0.14   &   0.1  \\
                     &SkM$^\ast$& GS  & -750.32  & -745.23   &   5.09   &  0.23   &  28.1  \\ 
                       &  SkP   & GS  & -742.94  & -735.45   &   7.49   &  0.17   &  51.2 \\
                       &        & IS  & -742.79  & -734.71   &   8.08   &  0.13   &   0.2 \\
\hline
$^{96}{\textrm {Ge}}$  &  Gogny & GS  & -728.29  & -718.33   &   9.96   &  0.25   &  33.2  \\
                       &  SLy6  & GS  & -736.79  & -735.64   &   1.15   &  0.24   &  38.1 \\
                       &  SkI3  & GS  & -747.95  & -747.95   &   0.00   &  0.25   &  40.2 \\ 
                       &  SkT6  & GS  & -745.96  & -743.98   &   1.98   &  0.23   &  33.8  \\
                     &SkM$^\ast$& GS  & -754.86  & -749.99   &   4.86   &  0.26   & 26.1  \\
                       &  SkP   & GS  & -745.94  & -739.09   &   6.84   &  0.20   & 35.8  \\
\hline
$^{98}{\textrm {Ge}}$  &  Gogny & GS  & -729.64  & -719.63   &  10.01   &  0.26   &  31.0  \\
                       &  SLy6  & GS  & -738.75  & -737.47   &   1.28   &  0.27   &  31.1 \\
                       &        & IS  & -738.57  & -736.13   &   2.44   &  0.21   &  58.8 \\
                       &        & IS  & -737.46  & -733.99   &   3.47   &  0.28   &   1.9 \\   
                       &  SkI3  & GS  & -751.62  & -749.92   &   1.70   &  0.22   &  60.0 \\ 
                       &  SkT6  & GS  & -748.78  & -747.12   &   1.66   &  0.26   &  26.9  \\
                       &        & IS  & -748.57  & -746.00   &   2.58   &  0.25   &   0.0  \\
                     &SkM$^\ast$& GS  & -759.01  & -754.82   &   4.19   &  0.27   &  25.2 \\ 
                       &  SkP   & GS  & -748.52  & -742.38   &   6.15   &  0.22   &  28.6 \\
\hline
$^{100}{\textrm {Ge}}$ &  Gogny & GS  & -730.39  & -720.45   &   9.94   &  0.27   &  29.2  \\
                       &  SLy6  & GS  & -740.38  & -740.38   &   0.00   &  0.30   &  26.6 \\
                       &  SkI3  & GS  & -754.00  & -753.62   &   0.38   &  0.24   &  45.6 \\ 
                       &        & IS  & -753.03  & -748.82   &   4.21   &  0.09   &  60.0 \\ 
                       &  SkT6  & GS  & -751.21  & -750.26   &   0.94   &  0.28   &  24.6  \\
                     &SkM$^\ast$& GS  & -762.68  & -759.16   &   3.52   &  0.27   &  25.2  \\
                       &  SkP   & GS  & -750.68  & -745.18   &   5.50   &  0.23   &  24.9  \\ 
\hline
$^{102}{\textrm {Ge}}$ &  Gogny & GS  & -730.52  & -720.34   &  10.18   &  0.26   &  28.1  \\
                       &  SLy6  & GS  & -741.30  & -741.29   &   0.01   &  0.27   &  31.5 \\
                       &        & IS  & -739.07  & -733.34   &   5.73   &  0.00   &   0.1 \\
                       &  SkI3  & GS  & -757.30  & -751.74   &   5.56   &  0.00   &   1.1 \\
                       &        & IS  & -756.30  & -755.13   &   1.17   &  0.21   &  33.8 \\ 
                       &  SkT6  & GS  & -753.15  & -752.52   &   0.63   &  0.25   &  21.3 \\
                     &SkM$^\ast$& GS  & -765.84  & -762.55   &   3.30   &  0.26   &  24.2 \\ 
                       &  SkP   & GS  & -752.39  & -747.36   &   5.03   &  0.23   &  22.6 \\
\end{tabular}
\end{ruledtabular}
\end{table}

\begin{table}
\caption{\label{tab4}(continued) as table~\ref{tab3}.}
\begin{ruledtabular}
\begin{tabular}{cccccccc}
nucleus & force & state &$E_B$[MeV] & $E_{hf}$ & $E_{pair}$ & $\beta$ & $\gamma$ \\
\hline
$^{104}{\textrm {Ge}}$ &  Gogny & GS  & -730.09  & -720.17   &   9.92   &  0.25   &  26.4  \\
                       &  SLy6  & GS  & -742.02  & -741.46   &   0.56   &  0.23   &  23.8 \\
                       &  SkI3  & GS  & -758.71  & -758.70   &   0.01   &  0.24   &  26.0 \\
                       &        & IS  & -758.61  & -753.52   &   5.09   &  0.06   &  56.2 \\
                       &  SkT6  & GS  & -754.43  & -753.39   &   1.04   &  0.23   &  22.0 \\
                     &SkM$^\ast$& GS  & -768.51  & -765.80   &   2.70   &  0.25   &  23.5 \\ 
                       &  SkP   & GS  & -753.45  & -749.22   &   4.24   &  0.23   &  21.3 \\
\hline
$^{106}{\textrm {Ge}}$ &  Gogny & GS  & -729.10  & -719.78   &   9.32   &  0.23   &  25.2 \\
                       &  SLy6  & GS  & -742.22  & -742.11   &   0.11   &  0.24   &  24.8 \\
                       &  SkI3  & GS  & -760.98  & -757.24   &   3.74   &  0.11   &   0.0 \\ 
                       &        & IS  & -760.79  & -760.76   &   0.02   &  0.19   &  24.5 \\ 
                       &  SkT6  & GS  & -755.31  & -754.93   &   0.38   &  0.23   &  22.0 \\
                       &        & IS  & -755.30  & -752.98   &   2.32   &  0.19   &   7.7 \\
                     &SkM$^\ast$& GS  & -770.60  & -768.31   &   2.29   &  0.23   &  22.3 \\
                       &  SkP   & GS  & -754.07  & -750.25   &   3.82   &  0.22   &  18.4 \\
\hline
$^{108}{\textrm {Ge}}$ &  Gogny & GS  & -727.47  & -718.11   &   9.36   &  0.21   &  24.1  \\
                       &  SLy6  & GS  & -742.21  & -742.21   &   0.00   &  0.19   &  18.9 \\
                       &  SkI3  & GS  & -763.39  & -761.50   &   1.89   &  0.14   &   0.0 \\ 
                       &        & IS  & -762.10  & -762.09   &   0.01   &  0.20   &  25.4 \\ 
                       &  SkT6  & GS  & -756.02  & -754.91   &   1.11   &  0.18   &   1.0 \\
                     &SkM$^\ast$& GS  & -772.28  & -769.50   &   2.77   &  0.21   &  19.4 \\
                       &  SkP   & GS  & -754.39  & -750.29   &   4.09   &  0.21   &   0.3 \\ 
\end{tabular}
\end{ruledtabular}
\end{table}

The potential energy landscape, deformation energy and shapes are
determined by an interplay of mean-field effects (shell structure) and
pairing. The pairing energies are shown as complementary information
in the lower panels of figure \ref{fig:shapes2}. 
{The Gogny force shows generally the largest pairing energies.
This is due}
%
%
to the quite
different treatment of pairing in both models. Our Skyrme calculations
use a rather sparse phase space for pairing, $\pm 5$ MeV about the
Fermi surface, while the Gogny calculations include a larger
space. 
{Moreover, the fitting of the pairing forces was done
differently. The Skyrme forces used the full odd-even staggering
as information while for the Gogny force, the staggering
was enhanced by about 20\% \cite{Gog80} to encounter for
spin polarization effects in these data \cite{Rut98a}.}  
But the overall scaling in pairing energy is not important because it
has negligible effect on the global observables due to a subtle balance
between mean-field energy and pairing contribution.  What we should
compare are the trends. And even here, we see noteworthy
differences. For all Skyrme forces, {particularly for SLy6 and
SkI3}, there is occasionally a breakdown of pairing which indicates
closeness to the phase transition because of the generally weaker
pairing.  A significant difference is also seen for the trends in the
region of very neutron-rich isotopes. The Gogny force shows a strong
increase while all Skyrme forces tend to shrink pairing
strength. This result, however, has to be taken with care. We are here
employing the pairing with a zero-range two-body force in Skyrme
HFBCS, but it is known that a density-dependent pairing force enhances
pairing in the regime of exotic nuclei \cite{Bender00,Dob01a}. The
relevant information for our purposes is that the predictions on
deformations, and particularly on triaxiality, are robust with respect
to quite different treatments of pairing.

We have corroborated this statement by studying the effect of varying
pairing strength for SLy6. We find that the overall deformation is
little influenced by varying the pairing strength, especially for the
nuclei in the neutron-rich region.  This means that pairing does not
overrule deformation effects dictated by nuclear shell
structure. This statement, though, has to be taken with a grain of salt.  There
are sometimes exceptions near sub-shell closures where a larger
sensitivity to pairing is observed.
Somewhat more sensitivity to pairing strength is seen concerning
triaxiality. That is not surprising as the energy gain from triaxial
deformation is smaller and can be more easily countered by
pairing. The general triaxial softness, however, persists in these
cases.

\section{\label{level4} Conclusion}

We have systematically investigated the properties of the Ge isotopic
chain from neutron number $N=26$ to 76 in the framework of Gogny HFB
and Skyrme HF plus BCS. The Gogny HFB equation, where
the finite-range Gogny interaction provides both particle-hole and
particle-particle correlations simultaneously, was solved in a
three-dimensional harmonic oscillator basis with $z$-simplex and
$\hat{S_y ^T}$ symmetries.  The coupled HF plus BCS equations, where
the density-dependent Skyrme force and a $\delta$-pairing interaction
were used to treat the mean-field and pairing correlations,
were solved in three dimensional coordinate space without any symmetry
restrictions. Three conclusions emerge:

First, both theoretical models predict that most of the Ge isotopes
have triaxial features.  The binding energies and the deformations
$\beta$ and $\gamma$ agree {in general} very well with the available
experiments. The Skyrme HF plus BCS calculations yield
shape-coexistent isomers with quite different shapes but minor energy
differences in many Ge isotopes. This may indicate $\gamma$-softness
rather than true isomeric states, although the height of the
barrier between the ground state and the isomer would yet have to be
checked.

Second, the {five} Skyrme parametrizations SLy6, SkI3,
{SkM$^*$, SkP}, and SkT6 were used for studying the effects of
{effective mass and spin-orbit interaction}, investigating general vs.\
specific properties in these isotopes. We found that the predictions
with Gogny D1S, Skyrme SLy6, and SkI3 are quite similar with only a
few exceptions.  
SkM$^*$ is still quite close while SkP and SkT6 with effective
mass one
predicted quite different nuclear shapes and prefer more spherical
deformation for the nuclei between the proton drip-line and the stable
region. Comparing the properties among the {Skyrme} forces, we
conclude that
{a mix of symmetry energy and shell effects determines the
extrapolation to large neutron excess. From the shell effects, the
effective mass is influential in all regions while the impact of the
(isovector) spin-orbit interaction increases with increasing neutron
number.}

Last, a variation of pairing strength with the SLy6 force has been
considered to study the effect of pairing on the properties of the
ground states and coexistent shape isomers. We found that the pairing
strength has little effect on binding energy and quadrupole
deformation. The nuclear triaxiality, however, is more sensitive to
pairing strength. Triaxiality is a subtle shell effect and may be
overruled by pairing.

\begin{acknowledgments}
Lu Guo acknowledges support from the Alexander von Humboldt Foundation.
We gratefully acknowledge support by the Frankfurt Center for
Scientific Computing.
The work was supported in part by the BMBF (contract 06 ER 124).
\end{acknowledgments}

\bibliography{triax_Ge}


\end{document}